# A Reduced-Order CFD Approach for Intermediate grade Coronary Arterial Clinical Parameter Assessment

Oeshee Roy[1], Priyanshu Ghosh[1], Sayan Karmakar[2] and Supratim Saha [3],*

[1] Mechanical Engineering Department, Meghnad Saha Institute of Technology, Kolkata, India,
[2] Department of Civil Engineering, Jadavpur University, Kolkata, India.
[3] Department of Mechanical Engineering, Indian Institute of Technology Madras, Chennai, India.
*supratim@alumni.iitm.ac.in

**ABSTRACT**

Coronary heart disease (CHD) remains a top reason of mortality worldwide. This study introduces a novel approach by integrating patient-specific Multi-slice CT scans into CAD models and employing a one-dimensional numerical framework to assess varying degrees of stenosis. The computational analysis encompasses the entire arterial tree, with a particular focus on stenosed coronary arteries modelled using an analytical equation. One-dimensional characteristic equations, utilizing forward and backward characteristic variables, are used to derive essential parameters such as area and velocity. A model based on resistance with reflection coefficient set to zero and realistic pressure waveform input is applied at the outflow and inflow respectively. Boundary conditions generated from the 1D model, capturing global characteristics, are subsequently used to simulate a 2D axisymmetric model, which captures local characteristics. The numerical solvers are validated against literature results, ensuring grid independence. Fractional Flow Reserve (FFR) and instantaneous wave-free ratio (iFR) are calculated using various non-Newtonian models across different severities for higher order model. Additionally, the role of lesion length in stenosed coronary arteries is investigated. Numerical simulations are performed over one cardiac cycle, covering both systole and diastole phases. The results demonstrate that FFR and iFR decrease with increasing stenosis severity. This method provides a reliable and non-invasive diagnostic tool for evaluating the functional severity of coronary artery stenosis in clinical settings, effectively capturing both global and local hemodynamic characteristics.

## 1. INTRODUCTION

Coronary heart disease (CHD) is a foremost global health concern, causing around 20 million deaths annually. In India, the CHD mortality rate rose from 17% (2001-2003) to 23% (2010-2013) [1]. Effective assessment methods for coronary artery disorders are essential for improving treatment strategies, particularly in severe cases requiring angioplasty or surgery. While mild CHD is generally treated with medication, many cases present a clinical challenge due to their intermediate severity, complicating treatment decisions. Traditional clinical assessments often rely on invasive techniques, which can be costly and unsuitable for managing medically treated stenosis [2]. Non-invasive alternatives, such as computed tomography angiography (CTA), provide 3D images of blood vessels and the heart, allowing for the estimation of clinical parameters like Fractional Flow Reserve (FFR). However, the computational demands of realistic stenosis modelling and blood vessel closure can be high. Previous studies have explored 3D numerical models for FFR estimation, but these are resource-intensive. In contrast, one-dimensional (1D) simulations present a more efficient approach while maintaining reasonable accuracy. Sherwin et al.[3] established key equations for blood flow in vessels, and Mynard and Nithiarasu [4] contributed relationships between systemic and coronary circulatory systems, which are integrated into a comprehensive 1D arterial network model in the present study. This study introduces a reduced-order computational fluid dynamics (CFD) model that utilizes patient-specific CT scan data to predict various clinical parameters, including FFR and instantaneous wave-free ratio (iFR). Comparative studies have shown the effectiveness of 1D versus 3D models in assessing hemodynamic parameters in coronary arteries [5,6]. Unlike previous work, which focused on 3D modelling, our method significantly reduces computational costs while maintaining accuracy. Additionally, various boundary conditions have been explored to compute hemodynamic parameters, as demonstrated in earlier studies [7,8]. This study aims to develop a reliable, non-invasive diagnostic tool for evaluating the functional severity of stenosed coronary artery by integrating patient-specific CT scan data into a 1D CFD model. This approach provides a quick, cost-effective alternative to traditional invasive techniques, enhancing clinical decision-making for CHD management. Addressing the critical parameter of lesion length is essential in our analysis, as longer lesions have been shown to significantly influence hemodynamic parameters, thereby exacerbating the severity of blockages [9,10]. In this work, we will focus on analyzing the impact of lesion length to better understand its role in the progression of arterial disease. We use non-Newtonian blood flow models, including the power-law, Carreau-Yasuda, and Casson models, which can better capture the complex rheological behaviour of blood [11]. The novelty of our approach lies in combining global and local hemodynamic characteristics through a 1D-2D

## 2. Methodology

### 2.1 Computational domain

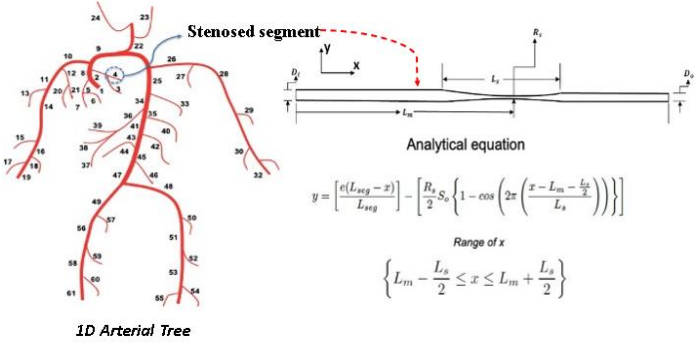

**Figure 1: Schematic illustration of the computational domain.**

The computational domain for the 1D numerical simulation covers the full arterial tree with 61 segments, where the stenosed section (segment 4) is located in the left part of the coronary tree, as shown in Figure 1. The stenosis is modelled using analytical equations that account for the tapering condition of the artery. In the 1D model, this stenosis is represented as a variation in cross-sectional area along the axial direction. The boundary conditions from the 1D model are then applied to a higher-order 2D axisymmetric model, where the same analytical equations are used to describe the stenosis. However, unlike the 1D model, the 2D axisymmetric model incorporates both axial and radial dimensions, providing a more detailed geometric representation of the artery. The severity of stenosis of patient-specific cases are 40, 50 and 70 which are of intermediate grade stenosis. The lesion length considered in the study are 1 cm and 3 cm. This hierarchical modelling approach allows for a comprehensive analysis, effectively capturing both the global characteristics of the entire arterial network and the local hemodynamic details within the stenosed region.

#### 2.1.1 1D Computational Model Description.

The vessel through which blood flows is modeled as a form closely resembling a cylindrical structure with a flexible wall. In the study by Sherwin et al. [3], the one-dimensional equations for mass along with momentum conservation are presented and shown in Equation (1) and Equation (2) respectively.

$$\frac{\partial A}{\partial t} + \frac{\partial (Au)}{\partial x} = 0 \qquad (1)$$

$$\frac{\partial u}{\partial t} + u\frac{\partial u}{\partial x} + \frac{1}{\rho}\frac{\partial p}{\partial x} - \frac{f}{\rho A} = 0 \qquad (2)$$

In this model, the cross-sectional area is represented by $A$, the average flow velocity by $u$, and $p$ denotes the pressure within the vessel. The blood density ($\rho$), is assumed to be 1060 kg/m$^3$, and $f$ indicates the frictional force. The friction term is modeled for steady and laminar flow conditions, consistent with Poiseuille flow [4, 5]. The system of equations is completed by introducing additional constraints that relate pressure to the vessel's cross-sectional area. The relationship between area and pressure is influenced by factors such as elasticity (wall behavior), wall thickness, and Poisson's ratio, as discussed in the literature [5, 10, 11].

$$p = p_{ext} + \beta (\sqrt{A} - \sqrt{A_0}) \qquad (3)$$

Where $p_{ext}$ represents the transmural pressure, $A_0$ is the cross-sectional area when the transmural pressure is zero (i.e. $p = p_{ext}$), and ($\beta$) denotes the material properties of the blood vessel.

$$A = \frac{(W_1 - W_2)^2}{1024}\left(\frac{\rho}{\beta}\right)^2 \qquad (4)$$

$$u = \frac{1}{2}(W_1 + W_2) \qquad (5)$$

The unknown variables, such as area ($A$) and velocity ($u$) are determined through the characteristics of forward and backward traveling waves, as outlined in Equations 4 and 5. In these equations, $W_1$ and $W_2$ correspond to the forward and backward wave characteristics, respectively. The numerical simulation is carried out using the Locally Conservative Galerkin (LCG) method [4, 10, 11].

### 2.1.2 Boundary conditions.

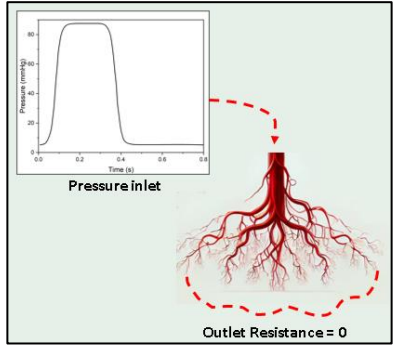

**Figure 2: 1D boundary condition implemented in arterial tree**

As shown in Figure 2, the schematic diagram illustrates the boundary conditions used in the 1D numerical model. A pressure boundary condition, created using a sigmoid function, is applied at the inlet of the arterial tree. At the outlet, a zero reflection coefficient is imposed. Further details about the boundary condition prescription using characteristic variables are available in the literature [4].

### 2.1.3 Grid independence study

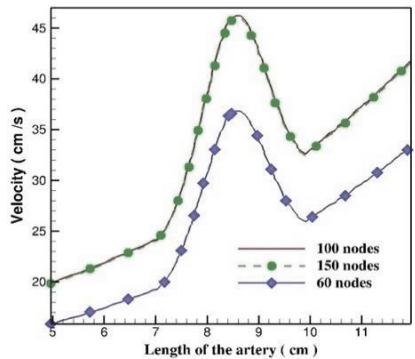

**Figure 3: 1D grid independent study of axial velocity**

For a given severity scenario, numerical analysis is conducted using three different mesh sizes: 50, 100, and 150 grid points. The magnitude of velocity in the artery's axial direction is calculated and compared across these mesh sizes, as depicted in Figure 3. The mesh sizes of 100 and 150 grid points show comparable trends. Consequently, the computational model employs 100 grid points per segment of arterial tree.

### 2.1.4 Validation study

The simulated results, illustrated in Figure 4, are compared with the numerical findings reported by Low et al. [12]. The right carotid artery was selected for validation with literature. The waveform patterns of both flow and pressure demonstrate good agreement with the literature results.

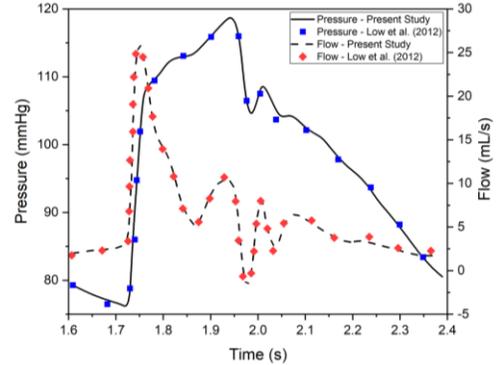

**Figure 4: Pressure and flow waveform comparison with literature results for Right Carotid artery [12]**

### 2.2.1 2D Numerical details

Assuming that flow of blood through the arterial vessel is incompressible and laminar, the governing equations for mass and momentum conservation are as follows.:

$$\frac{\partial \rho}{\partial t} + \nabla \cdot (\rho V) = 0$$

$$\rho \frac{\partial V}{\partial t} + \rho (V.\nabla)V = \nabla . [-pI + \mathbf{T}] + F$$

$$Where,\ T = \mu_{app} (\nabla.V + (\nabla.V)^T)$$

The Blood is modelled as Newtonian and non-Newtonian models like Casson model power law and Carreau Yesuda, model. The apparent viscosity ($\mu_{app}$) is different according to the model considered. The coefficient are taken from literature [13].

### 2.2.2 Boundary Conditions.

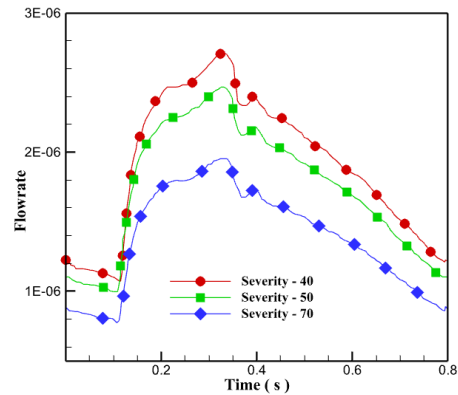

**Figure 5: 1D boundary condition generated for different severity which is given as inlet to higher order model**

The boundary conditions for 2D axisymmetric model are generated using 1D model for flowrate and pressure conditions. At inlet, flowrate is prescribed and at the outlet, pressure is imposed. The flowrate for different severity obtained from 1D model are shown in Figure 5. Table 2 shows the pressure conditons which are imposed at the outlet.

**Table 2: Pressure boundary condition generated from 1D model**

| Sr no | Severity ( % ) | Outlet pressure (mmHg) |
|---|---|---|
| 1 | 40 | 106.154 |
| 2 | 50 | 104.296 |
| 3 | 70 | 93.303 |

### 2.2.3 Grid independence study

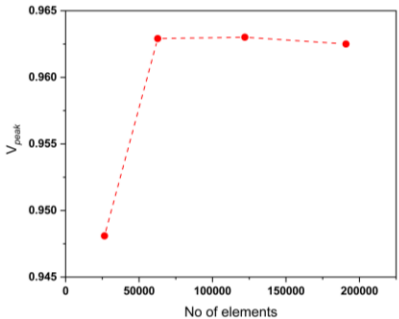

**Figure 6: Peak velocity variation with no of elements**

Grid independence analysis is executed for all stenosed cases, but Figure 6 illustrates it specifically for peak velocity ($V_{peak}$). The error is evaluated against the finest mesh used in the simulation. The simulation is carried out in the computational domain with approximately 200,000 elements to ensure a grid-independent solution.

### 2.2.4 Validation study

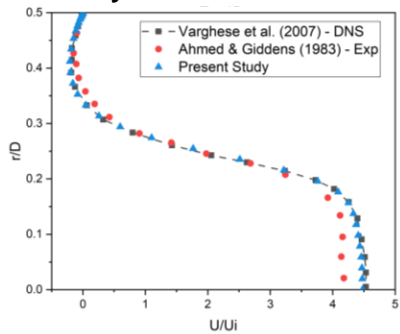

**Figure 7: Comparison of Velocity with literature result [14]**

The flow through a stenosed artery using 2D axisymmetric numerical model is validated against the results of Vargese et al. [14]. The Reynolds number (Re) is calculated using the characteristic length (D) and the reference velocity ($U_{avg}$) at the inlet, defined by the equation (Re = $\rho U_{avg} D/\mu$ ). The hydrodynamic parameters used in this simulation include a fluid density of 1000 kg/m$^3$ , a viscosity of 1 Pa·s , a reference velocity of 0.5 m/s and a diameter of 1 m. Both velocity and diameter are non-dimensionalized. At the inlet, a velocity having parabolic profile is applied, with a relative tangential velocity set to zero at the wall and a pressure of zero at the outlet [20,21]. The velocity profile at a specific location along the computational domain (1D), as shown in Figure 8, shows good agreement with results reported in the literature.

## 3. Result and discussion

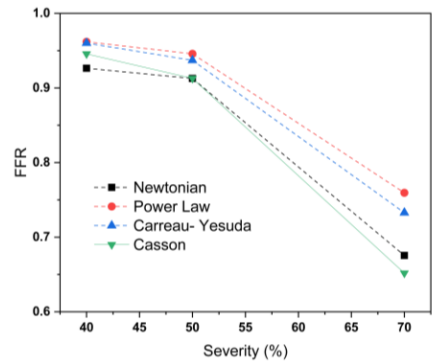

**Figure 8: 1D boundary condition given inlet to arterial tree**

The relationship between fractional flow reserve (FFR) and stenosis severity in coronary arteries is crucial for understanding blood flow dynamics and guiding clinical decisions. As stenosis severity increases, FFR values decrease, indicating greater flow impedance. This trend aligns with existing literature, where FFR values above 0.80 are generally associated with non-ischemic conditions, while values below 0.75 suggest significant ischemia-causing stenosis [15]. Notably, FFR values between 0.76 and 0.80 are often considered in a "grey zone," where clinical judgment is essential for determining the need for intervention, especially in cases of intermediate-grade stenosis [16]. In intermediate-grade stenosis, typically defined as 40% to 70% blockage, clinicians face a dilemma when deciding between clinical intervention and medical therapy. The FFR value decreases with increasing severity for all viscosity models and it is consistent with literature results [17,18]. At a stenosis severity of 40%, all models demonstrate high FFR values exceeding 0.92. Non-Newtonian models—such as the power-law, Carreau, and Casson—tend to predict slightly higher FFR values compared to the Newtonian model due to their shear-thinning properties. The power-law model, in particular, shows the least resistance and the highest FFR. As stenosis severity escalates to 50%, the Newtonian model predicts an FFR of approximately 0.91, while non-Newtonian models maintain marginally higher values. This pattern continues at 70% severity, where FFR values drop significantly, with the Newtonian model predicting the lowest FFR at about 0.68. In contrast, non-Newtonian models still predict higher FFR values, with the power-law model at 0.76, followed by the Carreau and Casson models at 0.73 and 0.65, respectively. These findings highlight the importance of model selection in clinical settings, particularly for intermediate-grade stenosis. While non-Newtonian models often yield higher FFR values, potentially indicating better flow conditions, there is a risk of overestimating FFR. Such overestimation could lead to underestimating the severity of blockages, adversely affecting treatment planning. Conversely, under-predicting FFR may result in overly aggressive interventions. Therefore, careful consideration is necessary when interpreting FFR results to ensure optimal patient outcomes. Current guidelines recommend the use of FFR as an adjunct to clinical decision-making, emphasizing that it should not replace comprehensive clinical assessment.

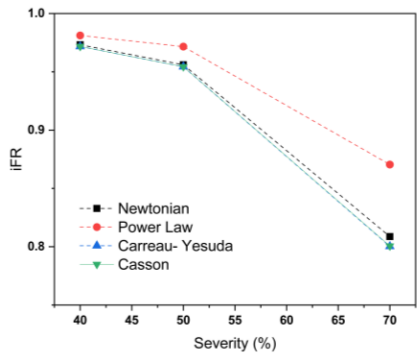

**Figure 9: Variation of iFR with severity**

The instantaneous wave-free ratio (iFR) provides a narrative similar to fractional flow reserve (FFR) in assessing coronary stenosis severity. The value of iFR decreases with increasing severity [11,19] as shown in Figure 9. At a stenosis of 40%, iFR values are high across all models, with the power-law model predicting the highest value at 0.981. As stenosis severity increases to 50% and 70%, iFR values correspondingly decrease, with the Newtonian model consistently yielding the lowest values. The power-law model maintains the highest iFR values at each severity level, highlighting its potential advantages in predicting blood flow dynamics. The gray zone for iFR, typically between 0.86 and 0.93, represents a range where the diagnostic certainty is lower. Clinical judgment is necessary to determine the appropriate course of action.

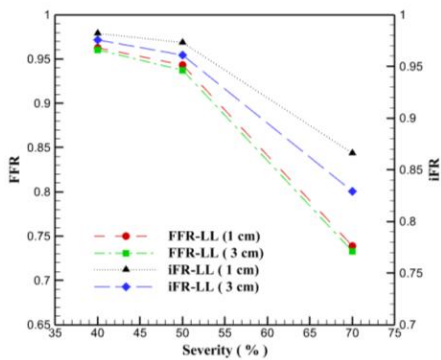

**Figure 10: FFR and iFR variation with severity for different lesion length**

Accurate diagnostic parameters, such as those provided by non-Newtonian models, are essential for evaluating coronary artery disease and guiding decisions between clinical intervention and medical therapy. This is particularly critical in cases of intermediate-grade stenosis, where the decision-making process can significantly impact patient outcomes. Figure 10 illustrates the comparison of FFR and iFR values for coronary artery lesions with varying severities and lengths (1 cm and 3 cm). These results highlight that both FFR and iFR measurements consistently show more severe blood flow obstruction as lesion length increases. This underscores the importance of considering lesion length in clinical evaluations of arterial blockages, as it significantly impacts the severity of blood flow obstruction and subsequent clinical decision-making . Accurate assessment using FFR and iFR can guide interventions by providing detailed insights into the hemodynamic impact of lesion length, aiding in more tailored and effective treatment strategies.

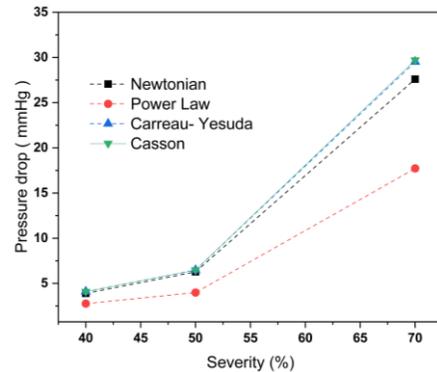

**Figure 11: 1D boundary condition given inlet to arterial tree**

Figure 11 demonstrate that as stenosis severity increases, the pressure drop also increases across all models. The Newtonian, Carreau-Yasuda, and Casson models display a similar pattern, with a steep rise in pressure drop at higher severity levels. The power-law model shows a more gradual increase in pressure drop, consistently indicating less resistance compared to the other models as severity worsens.

## CONCLUSION

This study introduces a novel approach by integrating patient-specific Multi-slice CT scans into CAD models and employing a one-dimensional numerical framework to assess coronary artery stenosis. The model captures both global and local hemodynamics using characteristic equations and a resistance model. Key findings include a decrease in Fractional Flow Reserve (FFR) and Instantaneous Wave-Free Ratio (iFR) values with increasing stenosis severity, indicating higher blood flow impedance. The study also compares different blood flow models—Newtonian, power-law, Carreau, and Casson—demonstrating that non-Newtonian models predict higher FFR and iFR values due to shear-thinning properties. Among these, the power-law model consistently shows the highest FFR and iFR values, indicating the least resistance, while the Carreau and Casson models yield slightly lower values than the Newtonian model. Additionally, the analysis emphasizes the importance of lesion length, as longer lesions (3 cm) result in lower FFR and iFR values compared to shorter lesions (1 cm). This underscores the significance of incorporating lesion length and non-Newtonian blood properties into computational models for more accurate, non-invasive assessments of coronary artery stenosis, ultimately enhancing clinical decision-making and patient outcomes.

**ACKNOWLEDGEMENTS**

The corresponding author acknowledges the PMRF scheme and funding (SB22230230MEPMRF008509) for their support in enhancing knowledge dissemination among students in private engineering colleges across India. This initiative plays a crucial role in bridging research gaps, fostering an environment that encourages young minds to actively participate in research activities.